\documentstyle[prd,aps]{revtex}

\newcommand{\diag}{\mathop{\rm diag}\nolimits}
\newcommand{\e}{\mathop{\rm e}\nolimits}
\newcommand{\ts}{\textstyle}
\newcommand{\ds}{\displaystyle}
\newcommand{\ints}{\int\limits}
\newcommand{\nn}{\nonumber\\}
\newcommand{\eps}{\varepsilon}
\newcommand{\Det}{\mathop{\rm Det}\nolimits}
\title{ \bf Chromomagnetic Catalysis of Color Superconductivity
and Dimensional Reduction}
\author{D.~Ebert $^{1,2}$, K.~G.~Klimenko $^{3}$, 
H.~Toki $^{1}$, and V.~Ch.~Zhukovsky $^{4}$}
\address{$^{1}$ Research
Center for Nuclear Physics (RCNP), Osaka University, Ibaraki,Osaka
567,Japan}
\address{$^{2}$ Institut f\"ur Physik,
Humboldt-Universit\"at zu Berlin, 10115 Berlin, Germany}
\address{$^{3}$ Institute
of High Energy Physics, 142284, Protvino, Moscow Region, Russia}
\address{$^{4}$ Faculty of
Physics, Department of Theoretical Physics, Moscow State University,
119899, Moscow, Russia}
\newcommand{\be}{\begin{equation}}
\newcommand{\ee}{\end{equation}}
\begin{document}
\large
\maketitle
\begin{abstract}
We consider diquark condensation in external chromomagnetic fields
at non--zero temperature. The general features
of this process are investigated for various
field configurations in relation to their symmetry properties and the
form
of the quark spectrum. According to the fields, there arises
dimensional
reduction by one or two units. In all cases there exists diquark
condensation
even at arbitrary weak quark attraction, confirming the idea about
universality
of this mechanism in a chromomagnetic field. Possible
influence of a nonzero chemical potential on the results obtained is
also discussed.

\end{abstract}
\renewcommand{\thefootnote}{\arabic{footnote}}
\setcounter{footnote}{0}
\setcounter{page}{1}
\section*{1. Introduction}
Nonperturbative effects in {\it QCD} at low energies
(large distances) can only be studied by approximate methods in the
framework of
various effective
models proposed. Among such nonperturbative effects are the existence
of the
{\it QCD} vacuum with
gluon and quark condensates \cite {1} and the hadronization process.
One of the
possibilities to
approximately describe the gluon condensate is to introduce
background color
fields of certain
configurations. One may, in particular, study the influence of
external
(background) color fields on
quarks \cite{alex}. In this case
it is possible to find expressions for the quark Green's functions
with exact consideration for the gauge field strength. This approach
enables one to make analytical calculations in order
to obtain estimates of various nonperturbative processes, such as
fermion condensate formation in constant non-Abelian fields \cite
{4},
thermodynamical stabilization of the vacuum state in an {\it SU(2)}
model of {\it QCD} with condensate fields \cite {5}, deep inelastic
hadron scattering influenced by gluon vacuum fields \cite{6} etc.

As is well known, the physics of light mesons can be described by
effective
four-fermion models such
as the Nambu--Jona--Lasinio (NJL) quark model, which was successfully
used to
implement the ideas of dynamical chiral
symmetry breaking ($D\chi SB$) and bosonization (see e.g. \cite {7}
and references therein; for a review of (2+1)-dimensional four-quark
effective models see \cite {echaya}). In particular, for a {\it
QCD}--motivated NJL--model with gluon condensate and finite
temperature, it was shown that a weak gluon condensate plays a
stabilizing role for the behavior of the constituent quark mass, the
quark condensate, meson masses and coupling constants for varying
temperature \cite {8}. The influence of temperature, chemical
potential \cite{zhetf}, and the external magnetic field \cite {9}
on the phase structure of various modifications of the
Nambu-Jona-Lasinio model was also discussed.

Moreover, it is in the framework of four--fermion models that a
constant magnetic field was shown \cite {10} to induce $D\chi SB$,
as well as the fermion mass generation, even under conditions when
the interaction between fermions is weak. Later, this phenomenon,
i.e., the effect of magnetic catalysis, was explained basing upon
the idea of effective reduction of space dimensionality in the
presence of a strong external magnetic field \cite {11} (see also
paper \cite {12} and references therein). It was also demonstrated
that a strong chromomagnetic (i.e., nonabelian) field catalyzes
$D\chi SB$ \cite {13}. As was shown in \cite {zheb}, this effect
can be understood in the framework of the dimensional reduction
mechanism as well, and it does not depend on the particular form
of the constant chromomagnetic field configuration.

Recently, the effect of diquark condensation and possible
color superconductivity (CSC),
has attracted much attention and
has been discussed in various publications
(see e.g.\cite {alf} -- \cite{klev}, and also the review
paper \cite{alf1} and references therein). One may expect that,
similar to the case of
the quark condensate, the process of
diquark condensation can
be catalyzed by intensive external
(vacuum) gauge fields. For a (2+1)- dimensional model, this was
recently discussed in \cite {ebklim}.

The purpose of the present paper
is to further investigate this possibility, now for a
(3+1)-dimensional model
including $(\bar q q)$-and $(qq)$-interactions,
for various external chromomagnetic fields like
non--abelian axial--sym\-met\-ric and rotational--sym\-met\-ric
ones,
as well as for abelian fields. In particular, we will show that in
all
cases, even for weak coupling of quarks, the diquark condensation
effect
induced by external chromomagnetic fields does exist and is related
to
an effective dimensional reduction. Moreover, we will find a simple
relation between symmetry properties of external fields, the
degeneracy
of quark energy spectra and the phenomenon of dimensional reduction.
The latter effect leads to a nonanalytic logarithmic dependence of
the
diquark condensate on the field strength in the strong field limit.
We shall also consider the effect of finite temperature and show
that in the strong field limit there exists a finite critical
temperature, at which a phase transition takes place and color
symmetry
is restored in both abelian and non--abelian models of the gluon
condensate. In particular, there arises
the BCS relation
$T_{C1}=C|\delta_0(0)|$ between the critical temperature and the zero
temperature diquark condensate $\delta_0(0)$, with a universal
constant
$C$ for different fields.
Finally, we shortly discuss the
influence of a nonzero chemical potential on the results obtained.
\section*{2. Quark and diquark condensates in external fields}

\subsection*{2.1 General definitions}

Let us consider
an
NJL model, which describes the
interaction of
flavored and colored
quarks $q_{i,\alpha}\,
(i=1,\dots,N_f,\,\alpha=1,\dots,N_c)$ with $N_f=2, N_C=3$ as numbers
of
flavors and colors, respectively
(for convenience, corresponding indices are
sometimes suppressed in what follows),
moving in an external chromomagnetic field.
The underlying quark Lagrangian
is chosen to contain four-quark interaction terms, which later on are
shown to be responsible for spontaneous breaking of both chiral
and color symmetries. Hence, two types of condensates
characterize the
ground state of the model: the quark
condensate $<\bar qq>$ (spontaneous breaking of chiral
symmetry), and the diquark condensate $<qq>$
(spontaneous breaking of color symmetry). Upon performing the usual
bosonization procedure
\cite{EbPerv},\cite {7}
and introducing meson and diquark fields $\sigma,\,\pi$
and $\Delta^b,\,\Delta^{*b}$, the four-quark terms are replaced by
Yukawa interactions of quarks with these fields, and the Lagrangian
takes the following form
(our notations refer to four--dimensional Euclidean space with
$it=x_4$)
\footnote{
We consider $\gamma-$matrices in the $4-$dimensional Euclidean space
with the metric tensor $g_{\mu\nu}=\diag (-1, -1, -1, -1)$, and the
relation between the Euclidean and Minkowski time
$x_{(E)}^0=ix_{(M)}^0
$: $ \gamma_{(E)}^0=i\gamma_{(M)}^0, \,
\gamma_{(E)}^k=\gamma_{(M)}^k.$ In what follows we denote the
Euclidean Dirac matrices as $\gamma_{\mu}$, suppressing the subscript
$(E).$ They have the following basic properties $\gamma_{\mu}^+= -
\gamma_{\mu},\,\{ \gamma_{\mu},\gamma_{\nu}\}=- 2\delta_{\mu\nu}.$
The charge conjugation operation for Dirac spinors is defined as $
\psi_c(x)\quad =\quad C\left(\psi(x)^+\right)^t $ with $
C\gamma_{\mu}^tC^{-1}=-\gamma_{\mu}.$ We choose the standard
representation for the Dirac matrices (see \cite{Rho}). The
$\gamma_5$
has the following properties:
$\{\gamma^{\mu},\gamma_5\}=0,\quad\gamma_5^+=\gamma_5^t=\gamma_5.$
Hence, one finds for the charge-conjugation matrix:
$C=\gamma^0\gamma^2,\quad C^+=C^{-1}=C^t=-C.$}:
\begin{eqnarray}
{\cal L} &=&-\bar q(i\gamma_\nu\nabla_\nu
+i\mu\gamma_0+\sigma+i\gamma^5\vec
\tau\vec\pi)q-\frac{1}{4G}(\sigma^2+\vec \pi^2)-
\frac{1}{4G_1}\Delta^{*b}\Delta^b-\nn
&-&\Delta^{*b}[iq^tC\varepsilon\epsilon^b\gamma^5 q]
-\Delta^b[i\bar q \varepsilon\epsilon^b\gamma^5 C\bar q^t].
\label{1}
\end{eqnarray}
Here $\mu$ is the chemical
potential, and $G, G_1$ are
(positive) four-quark coupling constants
(this becomes evident when integrating out the bosonic fields).
Furthermore,
$\nabla _\mu
=\partial_{\mu}-igA_{\mu}^a\lambda_a/2$ is the covariant
derivative of quark fields in the background field $F_{\mu
\nu }^a=\partial _\mu A_\nu ^a-\partial _\nu A_\mu ^a+gf_{abc}A_\mu
^bA_\nu ^c$ determined by the potentials $A_\mu ^a\left(
a=1,...,8\right)$, and
$\lambda_a/2$ are the generators of the
color $SU_c(3)$ group. Finally, $\vec \tau\equiv (\tau^{1},
\tau^{2},\tau^3)$ are Pauli
matrices in the flavor space, $\varepsilon$ and $\epsilon^b$ are
operators in the flavor and color spaces with matrix elements
$(\varepsilon)^{ik}\equiv\varepsilon^{ik}$,
$(\varepsilon^b)^{\alpha\beta}\equiv\varepsilon^{\alpha\beta b}$,
where $\varepsilon^{ik}$ and $\varepsilon^{\alpha\beta b}$ are
totally antisymmetric tensors, and $t$ denotes the transposition
operation.
 Clearly, the Lagrangian (\ref{1}) is
invariant under the color $SU_c(3)$ and
the chiral $SU(2)_L\times SU(2)_R$ groups.

In order to investigate
the possible
generation of quark and
diquark condensates in the framework of the initial model
(\ref{1}), let us introduce the partition function $Z$ of the
system
\begin{eqnarray}
Z &=&\int dqd\bar{q}d\sigma d\pi_i d\Delta^b
d\Delta^{*b} \exp
\left[\int d^4x{\cal L}\right].
\label{2}
\end{eqnarray}

Next, we shall evaluate the
functional integral over meson and diquark fields in (\ref{2})
by using the saddle point approximation,
neglecting field fluctuations around the
mean-field (classical) values $<\sigma>=\sigma_{0}$,
$<\pi>=\pi_{0}=0$
\footnote{The vanishing of the pion mean-field is here related to
the assumed parity conservation of the ground state.}
and $<\Delta^b>=\Delta^b_{0}$, $<\Delta^{*b}>=\Delta^{*b}_{0}$.
We then obtain the following gap equations
\begin{eqnarray}
-\frac {1}{2G}\sigma_{0}=<\bar{q} q>;~~~
-\frac{1}{4G_1}\Delta^{b}_0=<[iq^tC\varepsilon\epsilon^b\gamma^5 q]>
;~~~
-\frac{1}{4G_1}\Delta^{*b}_0=<[i\bar q \varepsilon\epsilon^b\gamma^5
C\bar q^t]>.
\label{delta1}
\end{eqnarray}
Within this approximation, we obtain the quark contribution to
the partition function
\begin{eqnarray}
Z_q=\exp W_{E}
=\int dqd\bar{q} \exp \left[\int d^4x {\cal L}_q\right],
\label{1k}
\end{eqnarray}
where
\begin{eqnarray}
{\cal L}_q =-\bar q(i\gamma_\nu\nabla_\nu
+i\mu\gamma_0+\sigma_0)q
-\Delta^{*b}_0[iq^tC\varepsilon\epsilon^b\gamma^5 q]
-\Delta^b_0[i\bar q \varepsilon\epsilon^b\gamma^5 C\bar q^t],
\end{eqnarray}
with $W_{E}$ being the Euclidean effective action, and ${\cal L}_q$
the
quark Lagrangian. It is evident that $ {\cal L}= {\cal L}_q +{\cal
L}_{scalar}$, where ${\cal L}_{scalar}$ is the lagrangian part of the
scalar meson and diquark.
Due to the fact that the partition function $Z_q$ is invariant
under the color gauge transformations, it is sufficient
in the following to study only the case with $\Delta^3_0\ne 0$
and $\Delta^{1,2}_0\equiv 0$. Hence, in the color superconducting
phase, where the diquark condensate is nonzero,
color symmetry  breaking
from $SU_c(3)$ symmetry down to $SU_c(2)$ takes place.
Furthermore, we also assume that the only
nonvanishing components of the potential are $A_{\mu}^a\ne
0,\,a=1,2,3$,
while others are equal to zero: $A_{\mu}^a=0,\,a=4,\dots,8$. This
implies
that only quarks of two colors $\alpha =1,2$ do interact with the
background field $A_{\mu}^a$, corresponding to the residual $SU_c(2)$
symmetry group of the vacuum. In this case,
the calculation of the quark
partition function (\ref{1k}) is greatly simplified,
and we have (for
more details see \cite{ebklim})
\begin{eqnarray}
Z_q=\Det_{(1)}(i\gamma\partial+\sigma_0+i\mu\gamma_0)
\cdot\Det^{1/2}_{(2)}\left [|\delta_0|^2
+(-i\gamma\nabla+\sigma_0+i\mu\gamma_0)
(i\gamma\nabla+\sigma_0+i\mu\gamma_0)\right ],
\label{intermediate}
\end{eqnarray}
where
$\delta_0=2\Delta^3_0$, and
indices (1) and (2) mean that determinants are calculated in the
one-dimensional (with $\alpha=3$) and in the
two-dimensional (with $\alpha=1,2$) subspaces of the color group,
respectively. In principle, the gap is complex and we have
two complex conjugated gap equations in (\ref{delta1}). However,
the partition function
is real and depends only on the module squared of the gap. Its phase
characterizes just the degeneracy of the vacuum and may be
set here equal to zero. For the general case, it is understood that
the gap equations
and the following determinants are expressed directly in terms of the
module $|\delta_{0}|$, i.e., $\delta_0 \rightarrow |\delta_0|$.

Let us
assume that the background field is constant and homogeneous,
$F_{\mu\nu}^a=const$. Then the Dirac equation $$ \left(i\gamma \nabla
+
\sigma_0\right)\psi=0 $$ for a quark with flavor $i$ has stationary
solutions $\psi_{k,i}$ with the energy spectrum $\varepsilon_{k,i},$
where $k$ stands for the quantum numbers of the quark in the
background
field. In this case we arrive at the following Euclidean effective
action: $$ W_E=\frac 12\int \frac{dp_4}{2\pi
}\{\sum_{k^{(0)},i,\kappa}\log
\left(p_4^2+(\varepsilon^{(0)}_{k^{(0)},i}-\kappa\mu)^2\right)+
$$
\begin{equation}
+\sum_{k,i,\kappa}\log
\left(p_4^2+|\delta_0|^2+(\varepsilon_{k,i}-\kappa\mu)^2\right)\}.
\label{3}
\end{equation}
Here, $\kappa=\pm1$ corresponds to charge conjugate contributions of
quarks, the first term in the sum corresponds to free quarks (not
interacting with the color
$SU_c(2)$ field)
with color $\alpha=3$ and the
spectrum
$\varepsilon^{(0)}_{k^{(0)},i}=\sqrt{\sigma_0^2+\vec {p}^2},$
and the second term corresponds to quarks with color indices
$\alpha=1,2$ (included in the quantum number $k$) and the spectrum
$\varepsilon_{k,i},$ moving in the background color field
$F_{\mu\nu}^a\,(a=1,2,3)$.

In the case of finite temperature $T= 1/
\beta >0$, the thermodynamic potential $\Omega=-W_E/(\beta L^{3})$
\cite{5} is obtained after substituting $p_4\rightarrow \frac{2\pi
}\beta (l+\frac 12), l=0,\pm 1,\pm 2,...$,
\begin{eqnarray}
\Omega&=&-\frac 1{\beta L^{3}}\sum^{N_f}_{i=1}\sum_{\kappa}
\sum^{l=+\infty}_{l=-
\infty}\biggl\{\sum_{k^{(0)}}\log\left[\left(\frac{2\pi(l+1/2)}{\beta
}\right)^2+
(\eps_{k^{(0)},i}-\kappa\mu)^2\right] \nn
&&+\sum_{k}\log\left[\left(\frac{2\pi(l+1/2)}{\beta}\right)^2+|\delta
_0|^2+
(\eps_{k,i}-\kappa\mu)^2\right]\biggr\}.
\end{eqnarray}

Next, let us consider
the proper time representation
\begin{eqnarray}
\Omega&=&\frac 1{\beta L^{3}}\sum^{l=+\infty}_{l=-\infty}
\sum^{N_f}_{i=1}\sum_{\kappa} \ints^{\infty}_{1/\Lambda^2}\frac{ds}s
\exp\biggl[-s\left(\frac{2\pi(l+1/2)}{\beta}\right)^2\biggr]\nn
&&\times\biggl\{\sum_{k^{(0)}}\exp[-s(\eps_{k^{(0),i}}-\kappa\mu)^2]+
\sum_{k}\exp[-s(|\delta_0|^2+(\eps_{k,i}-\kappa\mu)^2)]\biggr\},
\label{mu}
\end{eqnarray}
where $\Lambda $ is an ultraviolet cutoff ($\Lambda \gg
\sigma_0,\,|\delta_0|$). According to (\ref{2}) we then find for the
quark condensate
\begin{eqnarray} \langle \bar{q}q\rangle &=&
\frac{\ds\int\,d\bar qdq ~~ \bar{q}q \exp\left[\ds\int\,d^4x {\cal
L}_q \right]} {\ds\int\,d\bar{q}dq \exp \left[\ds\int\,d^4x {\cal
L}_q \right]}=-\frac{1}{ Z_q}\frac{\partial Z_q}{\partial
\sigma_0}=\frac {\partial \Omega}{\partial \sigma_0}, \label{5}
\end{eqnarray} which gives (for simplicity, we start with the
assumption that $\mu=0,$ and later return to the discussion of the
general case of $\mu \ne 0$)
\footnote {
It is well kown that for usual superconductors the
instability of the normal-fluid phase occurs in finite density
systems with $\mu \ne 0$. Notice that the possibility for vacuum CSC
at $\mu=0$ was recently studied in several papers
\cite{ebklim,chodos,van,ying}.
Since the chemical potential is a factor promoting the
appearance of CSC, we put it first equal to zero in order to get a
better
understanding of the role just played by the external chromomagnetic
field in the CSC formation. In this case the Fermi surface is
replaced by the ``zero energy surface'' $E=0$ of the system.
Due to the attraction between $\bar qq$- or $qq$-pairs, the
zero-energy
surface might become unstable and $\bar qq$- or
$qq$-pairs or both are produced depending on the values of $G,G_1$.
Then the true vacuum is the coherent superposition of the
$\bar qq$- and Cooper-pairs.}
 \begin{eqnarray} \langle \bar{q}q\rangle &=&
-\frac{\sigma_0}{L^{3}\sqrt{\pi}}\ints^{\infty}_{1/\Lambda^2}
\frac{\ds
ds}{\sqrt{s}}\left[1+2\sum^{\infty}_{l=1}(-1)^l
\e^{\ts-\frac{\beta^2l^2}{4s}}\right]\nn
&&\times\left(\sum_{k_{(0)},i}\e^{\ts-s\eps^2_{k_{(0)},i}}
+\sum_{k,i}\e^{\ts-s(\eps^2_{k,i}+|\delta_0|^2)}\right).
\label{6}
\end{eqnarray}
Here, the first term in the square brackets corresponds to the $T=0$
contribution, while the second term is the finite temperature
contribution ($T\ne 0$).

The
(scalar isoscalar)
diquark condensate can be obtained in a similar way
\begin{eqnarray}
\langle q q\rangle\equiv\langle iq^tC\varepsilon\epsilon^3\gamma^5
q\rangle &=& 2\frac
{\partial \Omega}{\partial \delta^*_0}.
\label{qq}
\end{eqnarray}
Hence, we have
\begin{eqnarray}
\langle qq\rangle &=&
-\frac{\delta_0}{L^{3}\sqrt{\pi}}\ints^{\infty}_{1/\Lambda^2}
\frac{\ds ds}{\sqrt{s}}\left[1+2\sum^{\infty}_{l=1}(-1)^l
\e^{\ts-\frac{\beta^2l^2}{4s}}\right]\nn &&\times \sum_{k,i}
\e^{\ts-s(\eps^2_{k,i}+|\delta_0|^2)}. \label{qqq}
\end{eqnarray}
Clearly, in the case of a vanishing external
field ($F_{\mu \nu }^a=0$), we have
$\varepsilon_k^2=\vec {p}^2+\sigma_0^2$. Then, at $T=0$ one
obtains for the quark condensate
\begin{eqnarray}
\langle\bar{q}q\rangle &=&-\frac{\ds
\sigma_0N_f}{\ds 2\pi^{2}}
\ints^{\infty}_{\frac1{\Lambda^2}}\frac{\ds ds}{\ds
s^{2}}\left(\e^{\ts-s \sigma_0^2}\,+\,2\e^{\ts-s(
\sigma_0^2+|\delta_0|^2)}\right),
\label{7}
\end{eqnarray}
and for the diquark condensate
\begin{eqnarray}
\langle q q\rangle &=&-\frac{\ds \delta_0N_f}{\ds \pi^{2}}
\ints^{\infty}_{\frac1{\Lambda^2}}\frac{\ds ds}{\ds
s^{2}}\e^{\ts-s( \sigma_0^2+|\delta_0|^2)}. \label{eight}
\end{eqnarray}

For subsequent discussion, this result can be easily generalized for
the case of a space-time of arbitrary dimensionality $D$:
\begin{eqnarray}
\langle q q\rangle &=&-\frac{\ds 4\delta_0N_f}{\ds
2^{D-2}\pi^{D/2}} \ints^{\infty}_{\frac1{\Lambda^2}}\frac{\ds
ds}{\ds s^{D/2}}\e^{\ts-s( \sigma_0^2+|\delta_0|^2)}.
\label{eighta}
\end{eqnarray}

In what follows, we shall analyze three special cases of external
chromomagnetic
fields.

{\bf Case $i$):}

Rotational--symmetric non--abelian chromomagnetic field
\begin{eqnarray}
A_1^1&=&A_2^2=A_3^3=\sqrt{\frac Hg},\quad H_i^a=\delta _i^aH
(i=1,2,3),
\label{8}
\end{eqnarray}
with all other components of $A_\mu ^a$ vanishing.

The energy spectrum has six branches, two of which correspond to
quarks that
do not interact with the chromomagnetic field
\begin{eqnarray}
\varepsilon _{1,2}^2&=&{\vec p}^2+\sigma_0^2,
\label{9}
\end{eqnarray}
and the other four are given as follows
\begin{eqnarray}
\varepsilon_{3,4}^2 &=& \sigma_0^2+(\sqrt{a} \pm \sqrt{{\vec
p}^2})^2,\nonumber\\
\varepsilon_{5,6}^2 &=& \sigma_0^2+(\sqrt{a} \pm \sqrt{4a +{\vec
p}^2})^2,
\label{10}
\end{eqnarray}
where $a=gH/4$.

{\bf Case $ii$):}

Axial--symmetric non--abelian chromomagnetic field
\begin{eqnarray}
A_1^1&=&A_2^2=\sqrt{\frac Hg}, H_i^a=\delta _3^a\delta _{i3}H,
\label{11}
\end{eqnarray}
with all other components of the potential vanishing.

The branches of the quark energy spectrum are besides (\ref{9}) as
follows
\begin{eqnarray}
\varepsilon_{3,4,5,6}^2&=&\sigma_0^2+2a \pm \sqrt{4a^2+4a
p^2_{\perp}}+p^2_3+p^2_{\perp}=\nn
&=&\sigma_0^2 +p^2_3+(\sqrt{ a +p^2_{\perp}}\pm \sqrt{a})^2.
\label{12}
\end{eqnarray}

{\bf Case $iii$):}

Abelian chromomagnetic field
\begin{eqnarray}
A_\mu ^a&=&\delta^a_3\delta_{\mu2}x_1H.
\label{13}
\end{eqnarray}
This time only two color degrees of freedom of quarks with "charges"
$\pm g/2$
interact with the external field. The energy spectrum of
quarks is now given by \begin{eqnarray}
\varepsilon_{3,4,5,6}^2&=&\eps^2_{n,\zeta,p_3}=gH(n+\frac12+
\frac{\zeta}2+
p^2_3+\sigma_0^2),
\label{14}
\end{eqnarray}
where $\zeta =\pm 1$ is the spin projection on the external field
direction, $p_3$ is the longitudinal component of the quark momentum
($-\infty
<p_3<\infty $),
\begin{eqnarray}
p_{\perp}^2&=&gH(n+\frac12)
\label{15}
\end{eqnarray}
is the transversal component squared of the quark momentum, and
$n=0,1,2,...$
is the Landau quantum number.
As can be seen from (\ref{6}) and (\ref{qqq}), the form of the
spectrum
is essential for the quark condensate formation. Using the above
three
expressions of energy spectra for field configurations $i$), $ii$)
and
$iii$), we shall next study the corresponding three types of quark
and
diquark condensates in the strong field limit.

\subsection*{2.2. Asymptotic estimates for strong fields
$gH\gg |\delta_0|^2, \sigma_0^2. $}

In this section, we consider the special cases of strong field
limits for the above configurations of background fields. Our goal
is here to demonstrate that the field is a catalyzing agent for
dynamical symmetry breaking, leading to  possible chiral
breaking and color breaking  (CSC) phases.  The fields are assumed
to be strong as compared to  the values of condensates that may be
rather small. In this sense, the expected values of fields
simulating the presence of a gluon condensate $gH=0.4-0.6\,\, {\rm
GeV}^2$ may be considered to be strong
\footnote{Notice that the value $gH=0.5\, GeV^2$ corresponds to a
chromomagnetic field of order $O(10^{19})$ Gauss which is much larger
than
the strong magnetic fields of order $O(10^{15})$ Gauss recently found
in
neutron stars, ``magnetars''\cite{ko}}.

{\bf Case $i$):}

According to (\ref{9}), (\ref{10}) we have
\begin{eqnarray}
\langle \bar{q}q\rangle
 =&&-\frac{\ds \sigma_0 N_f 4\pi a}{\ds \sqrt{\pi}(2\pi)^3}
\ints^{\infty}_{\frac{\ts a}{\ts \Lambda^2}}\frac{\ds dt}{\ds
\sqrt{t}}
\e^{\ts-\frac{tm_*^2}a}\ints^{\infty}_0\,dxx^2
\Bigl[2\e^{\ts-(x^2-\frac{|\delta_0|^2}{a})t}+\e^{\ts-t(1-x)^2}+\nn
&&+\e^{\ts-t(1+x)^2}
+\e^{\ts-(1+\sqrt{x^2+4})^2t}+\label{16aa} \\
&&+\e^{\ts-(1-\sqrt{x^2+4})^2t}\Bigr]\left(1+2\sum^{\infty}_{l=1}(-1)
^l \e^{\ts-
\frac{\beta^2l^2a}{4t}}
\right),\nonumber
\end{eqnarray}
and
\begin{eqnarray}
\langle qq \rangle &=& -\frac{\ds \delta_0 N_f 4\pi a}{\ds
\sqrt{\pi}(2\pi)^3} \ints^{\infty}_{\frac{\ts a}{\ts
\Lambda^2}}\frac{\ds dt}{\ds \sqrt{t}}
\e^{\ts-\frac{tm_*^2}a}\ints^{\infty}_0\,dxx^2
\Bigl[\e^{\ts-t(1-x)^2}+\nn &&+\e^{\ts-t(1+x)^2}
+\e^{\ts-(1+\sqrt{x^2+4})^2t}+\label{16a} \\
&&+\e^{\ts-(1-\sqrt{x^2+4})^2t}\Bigr]\left(1+2\sum^{\infty}_{l=1}(-1)
^l\e^{\ts-
\frac{\beta^2l^2a}{4t}}
\right),
\nonumber
\end{eqnarray}
where $m_*^2=|\delta_0|^2+\sigma_0^2.$
Taking the $T=0$ term in (\ref{16aa}), (\ref{16a}),
we see that the first term in the square brackets that corresponds to
the branch of the spectrum \[ \varepsilon _4^2=
\sigma_0^2+(\sqrt{a}-\sqrt{{\vec
p}^2})^2 \] plays the main role, when $h=gH/m_*^2=4a/m_*^2\gg 1$. In
this
case the following asymptotics are obtained
\begin{eqnarray}\langle\bar{q}q\rangle =-\frac{\sigma_0 N_f}{4\pi^2}
\left[3\Lambda^2-2m_*^2\log\frac{\Lambda^2}{m_*^2}-
\sigma_0^2\log\frac{\Lambda^2}{\sigma_0^2}+m_*^2\left( \frac
h2\log(C_{1}h)-
hI_1(\beta
m_*)\right)\right],
\label{17}
\end{eqnarray}
and
\begin{eqnarray}
\langle q q\rangle =-\frac{\delta_0 m_*^2N_f}{4\pi^2}
\left[2\left(\frac{\Lambda^2}{m_*^2}-\log\frac{\Lambda^2}{m_*^2}
\right) +\frac h2\log(C_{1}h)-hI_1(\beta m_*) \right]. \label{17a}
\end{eqnarray}
Here
\begin{eqnarray*}
I_1(\beta m_*)&=& -\sum^{\infty}_{l=1}(-1)^l
\ints^{\infty}_0\frac{dx}x \exp\left[-\left(x+\frac{l^2 m_*^2
\beta^2}{4x}\right)\right]=\nn &=&-2\sum^{\infty}_{l=1}(-1)^l
K_0(\beta
m_*l ), \end{eqnarray*} where $K_0(y)$ is the Macdonald`s function
and $C_1$ is a certain numerical constant.

It is well-known that
the order parameter of
$D\chi SB$ is the quark condensate which is the origin of dynamical
quark masses. At the
same time, color
superconductivity takes place, when the corresponding order
parameter,
the diquark condensate,
takes nonzero values.

The corresponding order parameters and underlying mechanisms of DSB
are
studied in this paper
for different chromomagnetic background fields (modelling the gluon
condensate) on the basis of an extended NJL model given in a
quark-meson
representation by the above Lagrangian (\ref{1}).
In the one--loop approximation
the gap equations (\ref{delta1})
can be rewritten
according to (\ref{17}), (\ref{17a}) in the form
\begin{equation}
\frac {\sigma_0}{2G}=\frac{\sigma_0N_f}{4\pi^2}\left[3\Lambda^2-
2m_*^2\log\frac{\Lambda^2}{m_*^2}-
\sigma_0^2\log\frac{\Lambda^2}{\sigma_0^2}+m_*^2\left(
\frac h2\log(C_{1}h)-hI_1(\beta m_*)\right)\right],
\label{si}
\end{equation}
and
\begin{equation}
\frac{\delta_0}{8G_1}=\frac{\delta_0 m_*^2N_f}{4\pi^2}
\left[2\left(\frac{\Lambda^2}{m_*^2}-\log\frac{\Lambda^2}{m_*^2}
\right) +\frac h2\log(C_{1}h)-hI_1(\beta m_*)
\right].\label{de}\end{equation} These equations have trivial
solutions $\sigma_0=0$ and $\delta_0=0,$ as well as nontrivial
ones. It is easily seen that the nontrivial condensates satisfy
the following gap equations:
\begin{eqnarray}
\Lambda ^2(\frac 1{\widetilde{g}}-1)&=& -\frac23 m_*^2
\log\frac{\Lambda^2}{m_*^2}-\frac13
\sigma_0^2 \log\frac{\Lambda^2}{\sigma_0^2}+
m_*^2 \frac h6 \log C_1h- h\frac{m_*^2}3I_1,
\label{19b}
\end{eqnarray}
where
$\widetilde{g}=\frac{\ts 3\Lambda^2 G}{\ts \pi^2 }$ ($N_f=2$),
and
\begin{eqnarray}
\Lambda ^2(\frac 1{\widetilde{g_1}}-1)&=& - m_*^2
\log\frac{\Lambda^2}{m_*^2}+
m_*^2 \frac h4 \log C_1h- h\frac{m_*^2}2I_1,
\label{19c}
\end{eqnarray}
where $\widetilde{g_1}=\frac{\ts 8 \Lambda ^2G_1}{\ts \pi^2 }$.
For $gH\log\frac{gH}{m_*^2}\gg m_*^2 \log\frac{\Lambda^2}{m_*^2}$
$(gH\ll \Lambda^2)$ we have solutions of (\ref{19b}), (\ref{19c})
even for weak coupling $\tilde g,~~ \tilde g_1\ll 1.$ It should be
noted, however, that in this weak coupling limiting case, the two
condensates may simultaneously take nontrivial values only for
$G=4G_1-y$, with $y$ being a small quantity, $y=O(G_1)$.
\footnote {Indeed, by using this relation and subtracting equation
(\ref{19b})
multiplied by a factor $\frac{3}2$ from (\ref{19c}), one gets a
simpler equation for $x=\frac{\Lambda^2}{\sigma_0^2}$ alone, which
admits a nontrivial solution for $0<y<\frac{4}3 \widetilde{g}G_1$.
This situation with a narrow region of coexisting phases in the plane
of coupling constants is, of course, due to our
approximation of very weak couplings that we  adopted in our
analytical estimates. A wider region of coexisting phases (phase IV)
was found
in our previous publication \cite{ebklim} in the d=3 model.
Preliminary results of numerical calculations show that, in the 4d
case considered in the present publication,  there also exists a
broader region of coexisting phases. It is this region that shrinks
to a line $G\simeq 4G_1$ in the limit of weak couplings. Detailed
results of our numerical investigations will be published
elsewhere.}
Otherwise they can exist separately.
In what follows, we investigate the case, when the
quark and diquark condensates are not simultaneously present. Then
the two phases are described by the formulas:
\begin{eqnarray}
\sigma_0(T)&=&\sqrt{C_1gH}\exp\left[-\frac{2\pi^2}{GgH}-I_1(\beta
\sigma_0(T))\right],\nn
\delta_0=0
\label{20}
\end{eqnarray}
for the case $G>4G_1$ or
\begin{eqnarray}
|\delta_0(T)|&=&\sqrt{C_1gH}\exp\left[-\frac{\pi^2}{2G_1gH}-I_1(\beta
|\delta_0(T)|)\right],\nn \sigma_0=0 \label{20a}
\end{eqnarray}
in the case $G<4G_1$ ($G=\frac{8}3 kG_1, k<\frac{3}2$).
In particular, for $T=0$,
\begin{eqnarray}
\sigma_0(0)&=&\sqrt{C_1gH}\exp\left(-\frac{2\pi^2}{GgH}\right).
\label{21}
\end{eqnarray}
The critical temperature $T_c$ can now be found from the condition
$\sigma_0(T_C)=0$,
which gives
(compare with \cite{zheb})
\begin{eqnarray}
T_C&=&\pi^{-1}\e^{\ts\gamma} \sigma_0(0)\simeq 0,5669\, \sigma_0(0).
\label{22}
\end{eqnarray}
 Similarly we obtain
\begin{eqnarray}
|\delta_0(0)|&=&\sqrt{C_1gH}\exp\left(-\frac{\pi^2}{2G_1gH}\right).
\label{21a}
\end{eqnarray}
Hence, for the critical temperature $T_{C1}$
of the phase transition, where $\delta_0(T_{C1})\to 0$, we
have the BCS relation
\begin{eqnarray}
T_{C1}&=&\pi^{-1}\e^{\ts\gamma} |\delta_0(0)|\simeq 0,5669\,
|\delta_0(0)|. \label{22a} \end{eqnarray}
For illustrations, let us quote a rough order of magnitude
estimate of $|\delta_0(0)|$ and $T_{C1}$. By
choosing, e.g. $\widetilde {g_1}<1$ and $gH<\Lambda^2$,
we obtain $|\delta_0(0)|<10\, MeV$ and $T_{C1}<5.7\, MeV$.
Let $\widetilde {g_1}=0.8,\, \Lambda^2=0.64 \,GeV^2$, then
$|\delta_0(0)|=7 \,MeV, T_{C1}=3.5 \,MeV$.

As is seen, the values of $T_C$ and $T_{C1}$ are here determined
by the values of corresponding condensates at $T=0$.
Notice that both condensates depend nonperturbatively on the
quantities $GgH, G_1gH$. Let us emphasize that the results
(\ref{17}), (\ref{17a}) with the logarithmic term $\frac h2\log h$
demonstrate the effect of dimensional reduction $D=3+1\rightarrow
D=1+1$. Indeed, integration of the main term in (\ref{16a}) gives
\begin{eqnarray}
\langle qq \rangle & \simeq & -\frac{\delta_0N_f a}{\pi^2}
\ints^{\infty}_{\frac1a} \frac{ds}s \e^{\ts-s m_*^2} \approx
-\frac{\delta_0 N_f a}{\pi^2} \log\frac{a}{m_*^2},
\label{23}
\end{eqnarray}
which, up to
a numerical factor, corresponds to (\ref{eighta}) with
$D=2$ and $\Lambda^2$ replaced by $a$.

{\bf Case $ii$):}

In this case we have for the diquark condensate
\begin{eqnarray}
\langle q q\rangle &=&-\frac {\delta_0 N_f}{(2\pi)^2}
\ints^{\infty}_{\frac1{\Lambda^2}} \frac{ds}s \e^{\ts-s m_*^2}
\ints^{\infty}_0\,dp_{\perp}p_{\perp}\times\nn &&\times\left[
\e^{\ts-s(\sqrt{a+p^2_{\perp}}-\sqrt{a})^2}+\right.\label{ii} \\
&&\left.+\e^{\ts-s(\sqrt{a+p^2_{\perp}}+\sqrt{a})^2}\right]
\left[1+2\sum^{\infty}_{l=1}(-1)^l
\e^{\ts-\frac{\beta^2l^2}{4s}}\right].
\nonumber
\end{eqnarray}

The gap equation for $h\gg 1$ now takes the form (when $\sigma_0=0$)
\begin{eqnarray}
\Lambda^2\left(\frac1{\tilde g_1}-1\right)&=&-|\delta_0|^2
\left(\log\frac{\Lambda^2}{|\delta_0|^2}-
\frac h2 -\frac{\sqrt{\pi h}}2I_2(\beta |\delta_0|)\right),
\label{24}
\end{eqnarray}
where
\begin{eqnarray*}
I_2(z)&=&\sum^{\infty}_{l=1}(-1)^l \ints^{\infty}_0\frac{dx}{x^{3/2}}
\exp\left[-\left(x+\frac{z^2l^2}{4x}\right)\right]=\\
&=&2\sqrt{\frac{\pi}{z}}\sum^{\infty}_{l=1}(-1)^l
\frac{\e^{\ts-\sqrt{z l}}}{\sqrt{l}}.
\end{eqnarray*}
It is convenient to rewrite (\ref{24}) in the form
\begin{eqnarray}
|\delta_0|&=&\Lambda
\exp\left[-\frac{\Lambda^2}{2|\delta_0|^2}\left(1- \frac1{\tilde
g_1}\right)- \frac h4-\frac{\sqrt{\pi h}}4I_2(\beta
|\delta_0|)\right].
\label{25}
\end{eqnarray}
The above solution is
valid, when the argument of the exponential function is negative.
Thus, for vanishing temperature $\beta \to \infty,$ we have the
condition
\[
\tilde g_1> \frac1{1+(gH/(2\Lambda^2))}.
\]
This demonstrates a possibility of color symmetry breaking in a non--
abelian chromomagnetic
field at $D=3+1$ even for $\tilde g_1<1$.

The dependence on $h$ in (\ref{24}) is found from the dominating
term in (\ref{ii}) arising from the branch
\[
\eps^2=\sigma_0^2+p^2_3+(\sqrt{p^2_{\perp}+a }-\sqrt{a })^2 .
\]
Then we have for $a \rightarrow\infty$
\[
\langle qq \rangle \sim \ints^{\infty}_{\frac1a}\frac{ds}s\e^{\ts-s
m_*^2} \ints^{\infty}_0\,dp_{\perp}p_{\perp} \e^{\ts-\frac{\ts
sp^4_{\perp}}{\ts 4a}} \sim \sqrt{a} \ints^{\infty}_{\frac1a}
\frac{ds}{s^{3/2}} \sim a \]
corresponding to (\ref{eighta}) with
$D=3$,
which demonstrates the $3+1 \rightarrow 2+1$ dimensional reduction in
this type of the field.

{\bf Case $iii$):}

For the abelian chromomagnetic field with the spectrum
(\ref{14})
we obtain
\begin{eqnarray}
\langle q q\rangle &=&-\frac {m_*^2\delta_0 N_f}{4 \pi^2}
\Bigl\{h\log\frac h{2\pi}+2\left(\frac{\Lambda^2}{m_*^2} -
\log\frac{\Lambda^2}{m_*^2}\right)- 2hI_1(\beta m_*)\Bigr\},
\label{26}
\end{eqnarray}
which is similar to (\ref{17a}), but differs by an overall factor 2
in
field-- dependent terms. This difference is simply due to the fact,
that the main term $h\log h$ is obtained from two colors in the
spectrum (\ref{14}), while in the non--abelian case only one branch
of
the spectrum contributes to (\ref{17}).
For $gH\log\left(\frac{gH}{|\delta_0|^2}\right) \gg
|\delta_0|^2
\log\left(\frac{\Lambda^2}{|\delta_0|^2}\right)$\,$(gH\ll
\Lambda^2)$ we obtain for $|\delta_0(T)|$, $|\delta_0(0)|$ and
$T_{C1}$ the same equations
(\ref{20a}), (\ref{21a}), (\ref{22a}) as in the
non--abelian case $i)$, but with the obvious replacements $C_1
\rightarrow 1/2\pi$ and $4 \pi^2\rightarrow 2\pi^2$ in the exponents.
The main logarithmic term in (\ref{26}) is obtained from the $n=0$,
$\zeta=-1$ contribution in the sum over quantum states in (\ref{qqq})
\begin{eqnarray*} \langle q q\rangle & \sim &
-\ints^{\infty}_{\frac1{\Lambda^2}}
\frac{ds}{\sqrt{s}}\ints^{+\infty}_{-\infty}\,dp_3
\sum^{\infty}_{n=0}
(2-\delta_{n0})\exp[-gHns-s |\delta_0|^2 -p^2_3 s]\sim \\
& \sim & -\ints^{\infty}_{1/gH}\frac{ds}s \e^{\ts-s |\delta_0|^2}
\approx -\log\frac{gH}{|\delta_0|^2}. \end{eqnarray*} Obviously, this
corresponds to
(\ref{eighta})
with $D=2$, which demonstrates the
dimensional
reduction
in this case $3+1\rightarrow 1+1$, similar to the non--abelian case
$i)$.
(The replacement $\Lambda^2 \rightarrow gH$ follows here from the
requirement
$1/\Lambda^2 \ll 1/gH \ll s$ for the integration region.)

As it is well
known, CSC is expected to appear at nonzero chemical potential (see,
e.g., \cite{rapp} and \cite{son}). Therefore, we now discuss a
possible influence of a finite chemical potential on our results. The
general case of arbitrary values of $\mu$ can only be considered by
numerical means. Neverthless, we can estimate its contribution to the
critical temperature analytically, when the gauge field is strong.
Let
us consider the most interesting case of a non-abelian field $i)$. As
we see from (\ref{mu}), including a finite $\mu$ can be made by the
replacement $\varepsilon ^2 \to \left(\varepsilon \pm \mu \right)^2$.
As was demonstrated above, the main contribution to the integral in
(\ref{16a}) comes from the branch in the energy spectrum
$\varepsilon _4^2= \sigma_0^2+(\sqrt{a}-\sqrt{{\vec p}^2})^2$. Hence,
for $\sigma_0 =0$ we have $\varepsilon ^2 = \left( \sqrt a - |\vec
p|\right)^2 \to \left( \sqrt a +\mu - |\vec p|\right)^2 $. Thus, to
account for the finite $\mu,$ we have to replace in the final
formulas
for $\delta_0$ and $T_{C1}$ in (\ref {21a}) and (\ref {22a}): $a^2
\to
\left(\sqrt a + \mu\right)^2.$ As a result we obtain for the critical
temperature the following estimate:
\begin{eqnarray}
T_{C1}&=&{\rm const}\left(\sqrt
{gH}+2\mu\right)\exp\left(-\frac{2\pi^2}{4G_1\left(\sqrt
{gH}+2\mu\right)^2}\right). \label{21b} \end{eqnarray} As follows
from the above estimate, the roles of $\mu$ and the vacuum field
$gH$ are complementary for the diquark condensate formation. It
should be mentioned that our result (\ref{21b}) reduces to formula
(6) of \cite {son}, when the chromomagnetic field vanishes.

\section*{3. Summary and discussions}

In the present paper the ability of external
chromomagnetic fields to induce dynamical symmetry breaking (DSB)
of chiral and color symmetry
was studied in the framework of the extended NJL model (1) with
attractive quark interactions
in  $qq$- and $\bar qq$-channels. Particular attention was paid
to the CSC generation. In order to understand the genuine
role of an external chromomagnetic field for the CSC phenomenon, we
have removed from our
consideration all other factors which might produce DSB. By this
reason, we have put the chemical potential
equal to zero, and considered  the weak coupling limit of the model
(temperature, on the other hand, is taken into account, since
this factor only promotes symmetry restoration and never induces
the DSB).

As was shown in our paper, the phenomenon of diquark condensation
does exist for various non--abelian chromomagnetic field
configurations even for the case of
weak coupling if $G<4G_1$ (if $G>4G_1$, the external
chromomagnetic field only catalizes the $D\chi SB$). This effect
is accompanied by an effective lowering of dimensionality in strong
chromomagnetic fields, where the number of reduced units of
dimensions depends on the concrete type of the field ---
a conclusion already made in the case of the $D\chi SB$ \cite{zheb}.
It should be mentioned that our result can be justified from the
general point of view.
Indeed, as the
$\sigma$ and $\Delta-$diquark fields appear in a typical combination,
the sigma meson and  diquark channel are
related by an approximate
Pauli--G\"ursey symmetry (see \cite{gurs}). If there appears a
catalysis phenomenon in the pure scalar sigma channel, it might be
expected
to appear also in the combination of sigma and diquark condensates.

The possibility for vacuum CSC at $\mu=0$ was also studied in the
framework
of random matrix models on the basis of general symmetry arguments
\cite{van}. There it was found a constraint on the coupling
constants in $qq$- and $\bar qq$-channels, at which the CSC
is forbidden. In terms of the NJL model (1) and at $H=0$ this
constraint
takes the form $G>8G_1/3$. We have, in particular, shown that the
external chromomagnetic
field modifies this constraint and reduces the region of
coupling constants, in which the CSC cannot occur. Indeed, in the
model (1) at $H\ne 0$ the range of prohibition for CSC is
$G>4G_1$ and is contained in the region $G>8G_1/3$.
Note that for NJL models based on the one-gluon exchange
approximation to
low energy QCD,
a simultaneous Fierz transformation into color-singlet $(\bar qq)$-
and antitriplet $(qq)$- channels yields the coupling relation
$G=8G_1$
(\cite{EbPerv,e}). Clearly, as our analysis shows, for such a case
the CSC cannot be produced at $\mu=0$ even at $H\ne 0$.

Finally, we remark that the chromomagnetic catalysis phenomenon for
the diquark
condensation is now under further examination, especially
with consideration for finite values of the chemical potential and
admitting various relations between coupling constants $G$ and $G_1$.
A preliminary
analysis of the interplay between various condensates
at $\mu \ne 0$, which is based on numerical
methods and generalizes our present results,
is given in \cite{ek} for the physically interesting case
$G=16G_1/3$.

\section*{Acknowledgements}

We wish to thank V.P. Gusynin, V.A. Miransky and Y. Nambu for
fruitful discussions.
One of the authors (D.E.) acknowledges the support provided to him
by the Ministry of Education and Science and Technology of Japan
(Monkasho) for his work at RCNP of Osaka University.
V.Ch.Zh. gratefully acknowledges the hospitality of
Prof. M.~Mueller-Preussker and his colleagues at the particle theory
group of the Humboldt University extended to him during his stay
there.
This work was partially supported by the Deutsche
Forschungsgemeinschaft under contract DFG 436 RUS 113/477/4.
 \end{document}